When Reflection and Transmission Amplitude Coefficients Exceed Unity: Evanescent Waves, Resonances, and Optical Modes

Raúl de la Fuente

Instituto de materiais da USC - iMATUS, Grupo de Nanomateriais, Fotónica e Materia Branda, Departamento de Física Aplicada, Universidade de Santiago de Compostela, E-15782, Santiago de Compostela, Spain

Abstract. The reflection and transmission of propagating harmonic waves in linear optical systems have been widely discussed in the literature and are generally well understood. In passive systems involving only propagating waves, energy conservation constrains measurable quantities such as reflectance and transmittance, as well as the corresponding amplitude coefficients. However, when evanescent or damped waves are present, the reflection and transmission coefficients of the field amplitude may exhibit atypical behaviour, with the real or imaginary parts being much larger than one or even unbounded. In this study, we analyse this phenomenon in several simple optical configurations, including dielectric–metal interfaces, surface-plasmon-resonance prism couplers, plasmonic microcavities, step-index slab waveguides and prism-coupled waveguides. We demonstrate that these large amplitude coefficients naturally arise from extending reflection and transmission coefficients to evanescent or damped waves. On the other hand, they are not directly associated with measurable reflectance or transmittance and, therefore, do not conflict with energy conservation. Instead, they provide a useful formal description of resonant field enhancement, modal excitation and coupling processes in plasmonic and dielectric optical systems.

Keywords: Fresnel coefficients, evanescent waves, resonant excitation

Fresnel coefficients are discussed in depth in many fundamental textbooks on optics. These coefficients express the relationship between the reflected and transmitted optical fields and the incident field, at an interface between two media with different optical properties. The Fresnel formalism provides a basis for analysing more complex systems involving one or more incoming and outgoing waves. Starting from the Fresnel formulas, generalised amplitude reflection and transmission coefficients relating these waves can be computed [1, 2]. These systems range from 1 × 1-port linear systems with one incoming and one outgoing wave (such as ideal mirrors or antireflection coatings) to multi-port systems (such as diffraction gratings or star couplers). Most systems have 2 × 2 port and are characterised by four amplitude coefficients associated with reflection and transmission from the left and right. All of them are combined to define the S, or scattering, matrix, which fully characterises the physical phenomenon [3–5]. While these coefficients relate the amplitudes of two optical fields (namely their electric or magnetic fields), reflectance and transmittance characterise energy or power ratios. The relationship between amplitude/energy coefficients is very simple in the case of reflection, where $R = |r|^2$. For the transmission coefficient, however, we must distinguish between transverse electric (TE) and transverse magnetic (TM) waves. For TE waves. $T = \Re[k_{on}]|t|^2/\Re[k_{in}]$ where $k_{in}(k_{on})$ is the normal component of the wavector for the input (ouput) medium. For TM waves. $T = \Re[n_o^* k_{on}/n_o]|t|^2/\Re[n_i^* k_{in}/n_i]$ where $n_i, n_o$ are the complex refractive index in the input and output media, respectively [6]. Both expressions for the transmittance coincide for transparent media, and $T = |t|^2$ when the input and output media are identical. The conservation of energy ($R + T \leq 1$) implies that the absolute value of the amplitude reflection coefficient is bounded by one, while the absolute value of the amplitude transmission coefficient is bounded by a factor of order unity depending on the value of the refractive indexes. In a related context, the modes of an optical waveguide can also be determined from these amplitude coefficients, following a simple procedure. First, the amplitude coefficients are defined for the specified optical system. Secondly, the incoming waves are set to zero. This leads to a non-trivial solution met at the poles of the amplitude coefficients that corresponds to the modal condition. This is a well-known fact in quantum mechanics, where bound states are related to the poles of the S-matrix. Consequently, an optical system admits modes when the corresponding amplitude coefficients have poles. For instance, Cardona [7] identified the pole of the reflection coefficient at a metal–dielectric interface as corresponding to a surface plasmon resonance, while the relationship between the poles of the amplitude reflection coefficients and the modes of multilayer waveguides was described in [1]. Furthermore, the input or output waves of a particular optical system need not be propagating harmonic waves but may instead be evanescent or damped waves. In this case, some amplitude coefficients can exceed one by a large amount. For instance, Simon [8] recognised in a study of surface plasmon excitation that the imaginary part of the relevant Fresnel reflection coefficient can attain very large values. In another study, Kalkak and Kumar (9) referred to energy propagation during the 'reflection' of evanescent waves, noting that the real part of the amplitude reflection coefficient can exceed unity without violating energy conservation. In a recent study of plasmonic microcavities, my co-authors and I demonstrated that the inner amplitude reflection coefficient of the mirrors can also be very large [10]. Despite the expected interest in this unusual behaviour, we believe that no comprehensive analysis of these atypical coefficients has yet been conducted. In the present study, we analyse several amplitude coefficients greater than unity, including unbounded ones, and demonstrate their relationship with resonances. However, this does not raise any issues regarding energy conservation because these coefficients are not directly related to measurable quantities such as reflectance or transmittance.

Before we begin our analysis, we would like to clarify the notation used. While the convention for defining the Fresnel coefficients for transverse electric (TE) waves is virtually universal, the same cannot be said for transverse magnetic (TM) waves. We adopt the definition as the ratio of the electric fields, and the sign convention for the electromagnetic fields is that of ref. [11].

Figure 1 shows the five configurations analysed in this work, ranging from a simple dielectric–metal planar interface to a step-index waveguide prism coupler. The first three configurations are related to surface plasmon resonances (SPRs), while the final two are related to step-index waveguides.

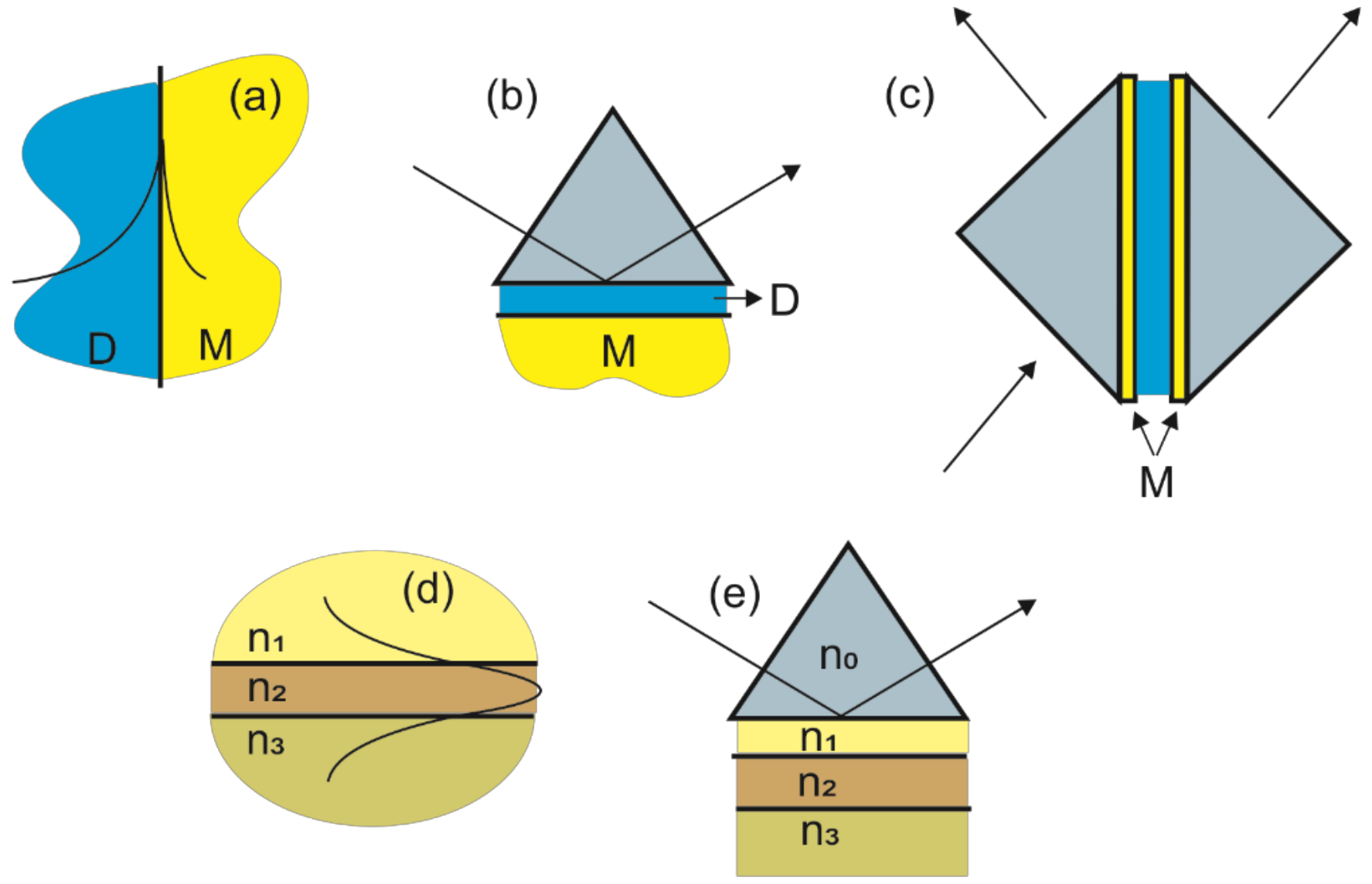

Fig. 1. (a) A dielectric-metal plane interface with a decaying wave. (b) The Otto configuration for surface plasmon resonance coupling. The Kretschmann configuration is obtained by interchanging the thin layer and the substrate. (c) A plasmonic microcavity with thin film metallic mirrors. (d) A planar step-index waveguide and (e) its prism coupling setting.

The first coefficient to be analysed is associated with surface plasmon resonance (SPR). Non-localized surface plasmon resonances are generated at the boundary between a dielectric medium with real permittivity $\varepsilon_d$ and a metallic medium with complex permittivity $\varepsilon_m = \varepsilon'_m + i\varepsilon''_m$, whose real part is negative, with $-\varepsilon'_m > \varepsilon''_m > 0$. The resonance condition is related to the Fresnel reflection coefficient for TM waves:

$$r_{dm} = -r_{md} = \frac{1+p}{1-p}; \quad p = -\frac{\varepsilon_d k_{mn}}{\varepsilon_m k_{dn}} \tag{1}$$

where $k_{jn} = \sqrt{\varepsilon_j k_0^2 - \beta^2}$, $k_0 = 2\pi/\lambda$ is the vacuum wavenumber and $\beta$ the component of the wavevector parallel to the boundary. Furthermore, $p$ is defined by choosing the appropriate branch of the square root, so that, $\Re[p] > 0$. Specifically, the mode condition corresponds to the pole of the Fresnel coefficient, $p = 1$, or:

$$\beta \equiv \beta_R = k_0 \sqrt{\frac{\varepsilon_m \varepsilon_d}{\varepsilon_m + \varepsilon_d}} \cong k_0 \sqrt{\frac{\varepsilon'_m \varepsilon_d}{\varepsilon'_m + \varepsilon_d}} \left[1 + i \frac{1}{2} \frac{\varepsilon''_m}{\varepsilon'_m} \frac{\varepsilon_d}{\varepsilon'_m + \varepsilon_d}\right] \tag{2}$$

where it has been assumed that $\varepsilon''_m \ll -\varepsilon'_m$. We shall make this assumption from now on. However, under normal circumstances $\beta$ is real and this condition is never met. Nevertheless, while for harmonic waves propagating in the dielectric ($\beta \le k_0\sqrt{\varepsilon_d}$) it follows that $|r_{dm}| \le 1$, when evanescent waves are present in the dielectric ($\beta > k_0\sqrt{\varepsilon_d}$), $|r_{dm}|$ can greatly exceed this value, approaching the modal value as $\varepsilon''_m \to 0$. We will consider this last regime. Writing $p = p' + ip''$, we get:

$$r_{dm} = -\frac{2(p'-1)}{(p'-1)^2 + p''^2} - 1 + i\frac{2p''}{(p'-1)^2 + p''^2}$$

(3)

To a good approximation, $p'$ and $p''$ are given by:

$$p' \cong -\frac{\varepsilon_d}{\varepsilon'_m}\sqrt{\frac{\beta^2 - \varepsilon'_m k_0^2}{\beta^2 - \varepsilon_d k_0^2}}$$

$$p'' \cong -p'\frac{\varepsilon''_m}{\varepsilon'_m}\left(1 + \frac{1}{2}\frac{\varepsilon'_m k_0^2}{\beta^2 - \varepsilon'_m k_0^2}\right)$$

(4)

The condition $p' = 1$ corresponds to $\beta = \Re[\beta_R]$ and then $r_{dm} \cong -1 + 4i\varepsilon'_m/\varepsilon''_m[\varepsilon'_m/(\varepsilon'_m - \varepsilon_d)]$. Meanwhile, the zero of the real part of $r_{dm}$ is achieved when $|p| = 1$, where $\beta^2 \cong \Re(\beta_R)^2\left[1 - \frac{1}{4}\frac{\Re(\beta_R)^2}{k_0^2}(\varepsilon_d - \varepsilon'_m)\left(\frac{\varepsilon''_m}{\varepsilon'_m}\right)^2\right]$, and the imaginary part exceeds its value for $p' = 1$ by a factor of order $(\varepsilon''_m/\varepsilon'_m)^2$. Furthermore, it corresponds, approximately, to the maximum of the imaginary part of $r_{dm}$.

Fig. 2 depicts the shape of $r_{dm}$ for an air-silver interface as a function of β and λ. The imaginary part displays a single peak whose value is significantly greater than one, as indicated by the above expression, whereas the real part shows a pronounced minimum and maximum around this peak, crossing zero near the maximum of the imaginary part. Meanwhile, the phase rapidly decreases from values close to π to values close to zero. As $-\varepsilon''_m/\varepsilon'_m$ decreases, the extremes of the real part of $r_{dm}$ increase, and the imaginary part approaches a delta-like function while the phase approaches a step function.

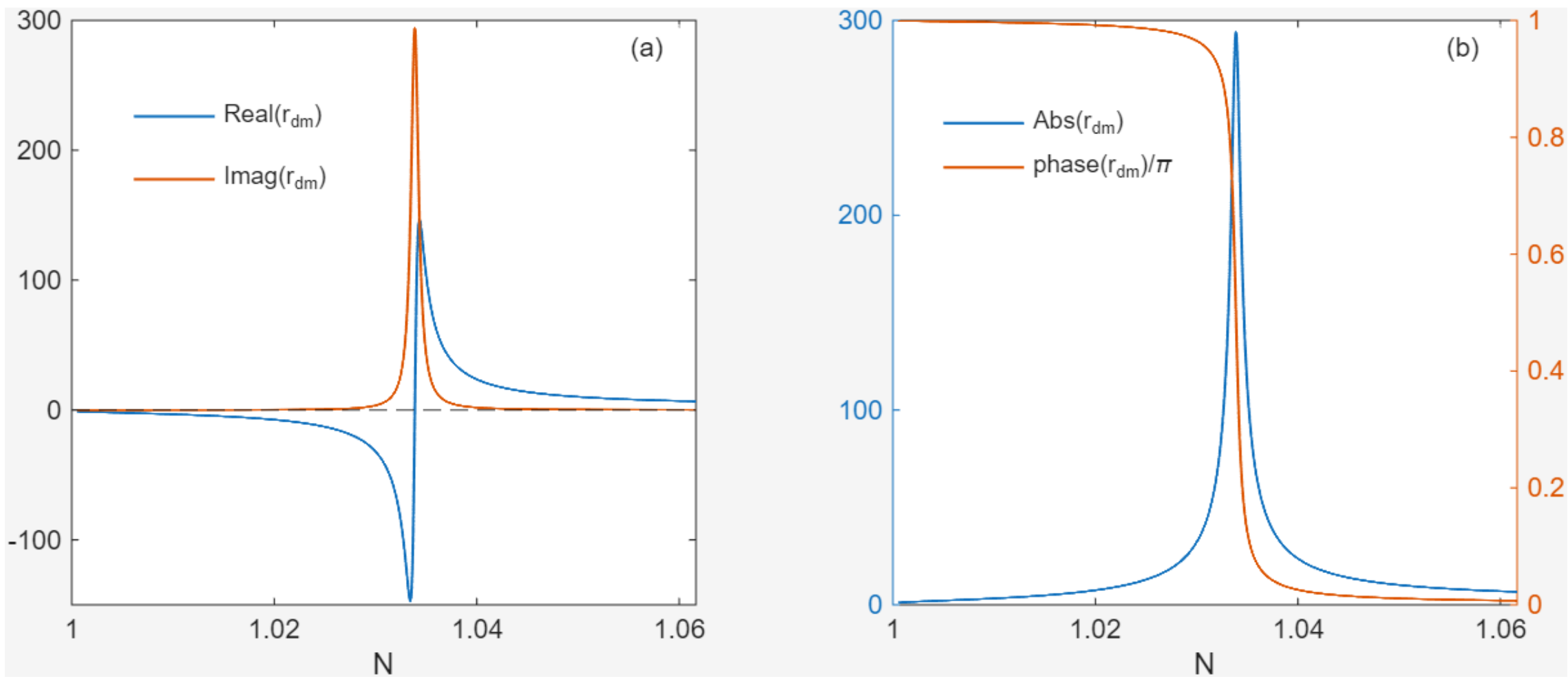

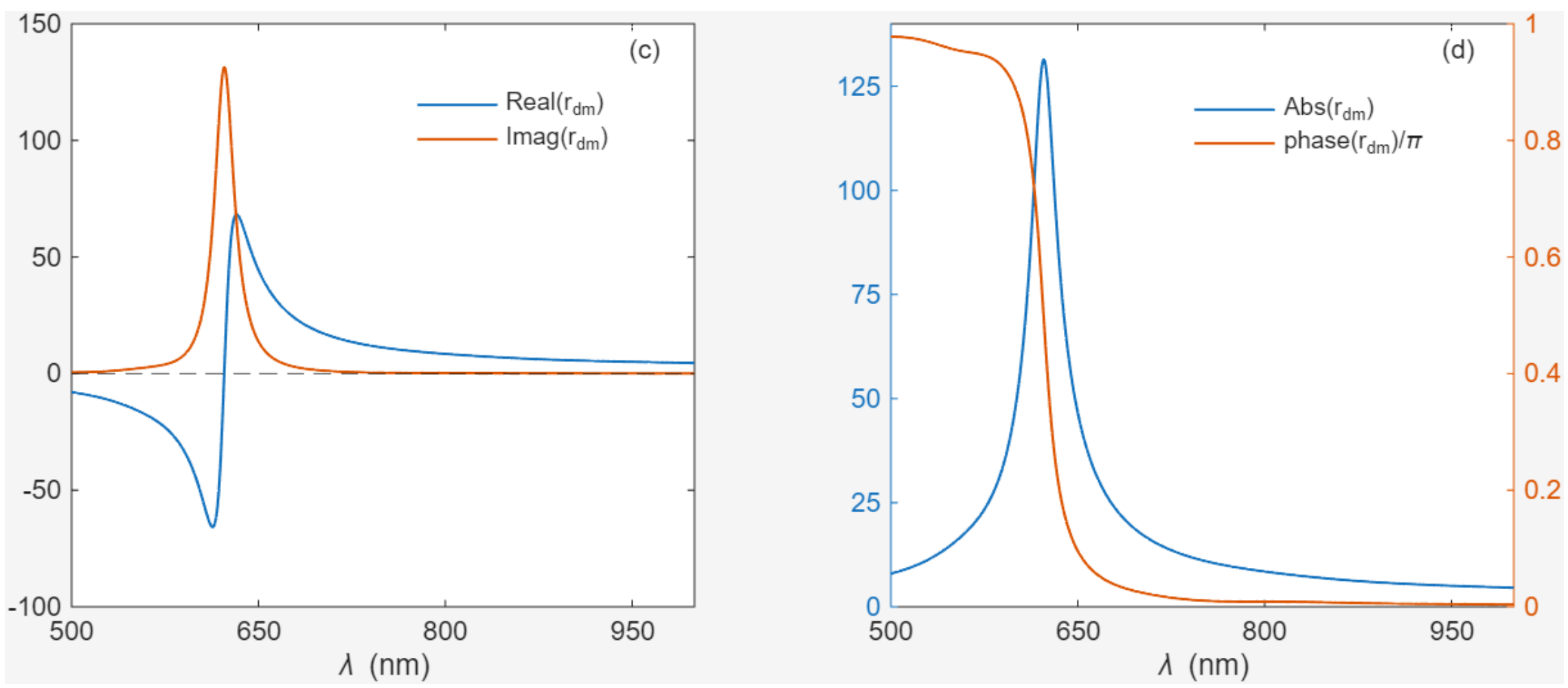

Fig. 2. Real and imaginary part of $r_{dm}$ (a), and absolute value and phase (b) as a function of the squared index $N^2 = (\beta/k_0)^2$ in a air – silver interface for λ = 800 nm ($\varepsilon_m = -31.021 + 0.40948i$, $\varepsilon_d$ =1.00055). Real and imaginary part of $r_{dm}$ (c), and absolute value and phase (d) as a function wavelength for $N = 0.68$. These last graphs are more irregular due to material dispersion.

The Kramers–Kronig relations, which are well known in the context of the complex permittivity of a medium, can also be applied to amplitude reflection coefficients if they are considered to be causal response functions of frequency. Under the usual analyticity conditions, the real and imaginary parts of the reflection coefficient are related through Hilbert-transform relations and are not independent. In many optical contexts, it is more convenient to write these relations in terms of the logarithm of the modulus of the reflection coefficient and its phase [12], linking the spectral variation of the phase to that of the amplitude. In the present case, however, the large values of the coefficient are more easily understood by parameterizing the real and imaginary parts of $r_{dm}$ in terms of the complex variable p, as shown in Eq. (3): $p''[\Re(r_{dm}) + 1] = (1 - p')\Im(r_{dm})$. This explicitly shows that the dispersive-like variation of the real part and the resonant peak of the imaginary part have the same origin; both arise from proximity to the pole at p = 1. Therefore, the behavior displayed in Fig. 2(c-d) is analogous to the typical resonance line shape of a causal optical response, featuring a rapidly varying real part and a strongly peaked imaginary part.

We have focused the analysis on the coefficient $r_{dm}$, but the boundary problem is characterized by the 2 x 2 scattering matrix:

$$S = \begin{pmatrix} r_{dm} & t_{md} \\ t_{dm} & r_{md} \end{pmatrix} \tag{5}$$

As indicated in Eq. (1) $r_{md} = -r_{dm}$. In addition, $t_{jk} = \sqrt{\varepsilon_j/\varepsilon_k}\,(1 + r_{jk}), j \neq k = d, m$. Consequently, all these Fresnel coefficients have real and imaginary part much larger than one.

Unfortunately, we cannot measure $r_{dm}$ directly. Nevertheless, this coefficient plays a prominent role in SPR prism-coupling configurations. There are two common configurations: the Kretschmann and Otto layouts (see Fig. 1(b)) [13]. In each case, there is a thin layer between a high-refractive-index medium and a bulk medium. In the Kretschmann configuration, the layer is metallic and the bulk medium is a low-refractive-index dielectric. In the Otto configuration, there is a thin dielectric layer of low refractive index in contact with a thick metal. To excite the plasmon resonance, the angle of incidence from the prism, $\theta$, must exceed the critical angle, $\theta_{cr} = \sin^{-1}\sqrt{\varepsilon_d/\varepsilon_p}$, where $\varepsilon_p$ is the prism permittivity.

In this case, there is one incoming wave and one outgoing wave, namely the reflected wave. This is a simple example of a 1 × 1-port linear system. Since the incident light comes from the prism, the scattering matrix reduces to $S = r_{pjk}$ with j = m, k = d in the Kretschmann layout and j = d, k = m in the Otto layout. This amplitude coefficient fully determines the SPR-coupling response. The expression for this amplitude reflection coefficient can be obtained recursively by applying Rouard's method [1], or equivalently by first calculating the transfer matrix of the system. One obtains:

$$r_{pjk} = r_{pj} + \frac{t_{pj}t_{jp}r_{jk}e^{i2k_{jn}d}}{1 - r_{jp}r_{jk}e^{i2k_{jn}d}} = \frac{-r_{jp+}r_{jk}e^{i2k_{jn}d}}{1 - r_{jp}r_{jk}e^{i2k_{jn}d}} \tag{6}$$

The expression of $r_{pjk}$ as the sum of two terms in eq. (6) has a simple interpretation. The first term, $r_{pj}$, corresponds to direct reflection, or equivalently to the contribution obtained in the limit $d \to \infty$. The second term modifies the reflection because light is transmitted through medium j and reaches medium k. In eq. (6), $r_{dm}$, or $r_{md}$, is the coefficient discussed previously, it describes the optical properties of the last dielectric-metal interface.

Both the Kretschmann and Otto configurations are usually analysed in relation to reflectance, $R = |r_{pjk}|^2 \leq 1$. Here, we consider the absorptance instead, because the resonance condition appears more clearly in this quantity. The analysis is simpler in the Otto layout since $|r_{pd}| = 1$ and $k_{dn} = ik''_{dn}$ for angles of incidence $\theta > \theta_{cr}$. In this case, we obtain:

$$A = 1 - R = \frac{\sin(\varphi_{dp})\sin(\varphi_{dm})}{\sinh^2[k''_{nn}(d - 0.5\ln|r_{dm}|/k''_{dn})] + \sin^2[(\varphi_{dp} + \varphi_{dm})/2]} \tag{7}$$

where d is the dielectric layer thickness. The amount of light absorbed depends strongly on two quantities:

$$D = 0.5\ln|r_{dm}|/k''_{dn}$$
$$A_0 = \frac{\sin(\varphi_{dp})\sin(\varphi_{dm})}{\sin^2[(\varphi_{dp} + \varphi_{dm})/2]} \tag{8}$$

They have the following meaning. For d = D, first, the argument of the hyperbolic sine function is zero and, second, $A = A_0$. Consequently, high absorptance, or equivalently low reflectance, is achieved provided that

$$D \to d$$
$$A_0 \to 1 \tag{9}$$

The last condition can be rewritten as $\varphi_{dp} \cong \varphi_{dm}$, for which $A_0$ reaches its limiting value. The first condition can be met for any angle above the critical one since $|r_{dm}| \geq 1$, yielding values of d that may exceed the wavelength as $|r_{dm}|$ approaches its maximum. By contrast, the second condition is only fulfilled for angles close to the peak of D and for grazing angles, see Fig. 3. Therefore, although there is a continuum of absorptance maxima, or reflectance minima, as d varies, only two regions of pronounced absorptance peaks and reflectance dips are actually obtained. This highlights the importance of the phase—and not just the amplitude—of the coefficient $r_{dm}$. Indeed, the phase jump that occurs alongside the peak of $r_{dm}$ guarantees a high absorptance near d = D. In Fig. 3 (c-d) we show absorption map as functions of $d$ and $\theta$, and $d$ and $\lambda$. We also plot the locus of $D = d$. This curve corresponds approximately to local maxima of absorbance; however, away from the resonances, $A_0$ is far from unity. It can also be seen that for $d > D$ there are additional local maxima of decreasing amplitude, located at angles lower than the main resonance angle, i.e. the resonance corresponding to $d \cong max(D)$).

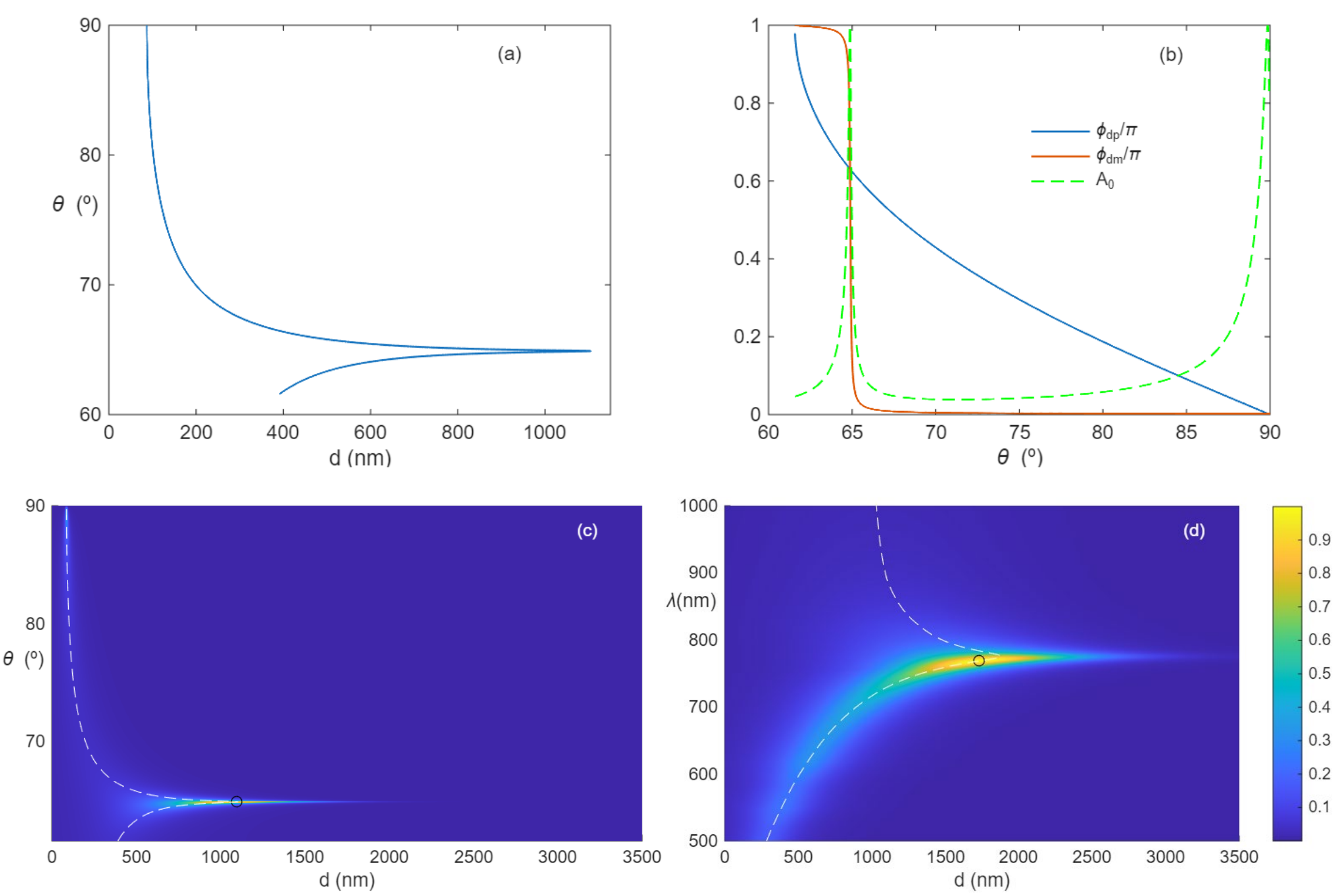

Fig.3. Plot of $d = D$ in the (d,θ) plane (a), and (b) phases $\varphi_{dp}, \varphi_{dm}$ and the amplitude factor $A_0$ as functions of the angle of incidence, for λ = 800 nm. The configuration is as follows: the prism is made of BK7 glass, the layer is air, and the metal is silver. Panels (c) and (d) show absorptances map in the $(d, \theta)$plane and in the $(d, \lambda)$plane, respectively. The white dash line corresponds to $d = D$ and the black circle to the absorptance maximum.

For the Kretschmann layout, the analysis is very similar, but with two main differences. First, $|r_{pd}|$ is not exactly equal to one, although it remains very close. Second, $k_{dn}$ is complex, with a small real part. In this case, only one pronounced resonance occurs around a specific metallic-layer thickness, close to the maximum of D. We emphasize that the fact that $|r_{dm}|$ exceeds unity makes it possible to achieve optimum values for the layer thickness in these SPR-coupling systems.

Despite the prominent role of $r_{dm}$ in these prism-coupling layouts, the relevant reflection coefficient $r_{pjk}$ has an absolute value smaller than one, so that the reflectance is well defined. This is not the case for the related coefficient $r_{jkp}$. Consider, for example, a plasmonic microcavity built using two symmetric inverted Kretschmann layouts [10,14], see Fig. 1(c). The scattering matrix of each mirror has as its elements the amplitude $r_{dmp}, r_{pmd}, t_{dmp}$ and $t_{pmd}$. The inner amplitude reflection coefficient of the mirror is

$$r_{dmp} = r_{dm} + \frac{t_{dm} t_{md} r_{mp} e^{i2k_{mn}w}}{1 - r_{mp} r_{md} d e^{i2k_{mn}w}} = \frac{-r_{md} + r_{mp} e^{i2k_{mn}w}}{1 - r_{mp} r_{md} e^{i2k_{mn}w}} \tag{10}$$

where $w$ is the mirrors thickness. In this case, the shape of $r_{dmp}$ closely follows that of $r_{dm}$ as shown in in Fig. 4. Depending on the cavity thickness, this structure exhibits one or two plasmonic resonances related to collective oscillations of free electrons at the two inner metal–dielectric boundaries of the cavity. These are usually referred to as coupled surface plasmon resonances. This resonances manifests as peaks of transmittance when d = D, with d the intracavity thickness and [10]:

$$D = 0.5 \ln\left|r_{dmp}\right| / k''_{dn} \tag{11}$$

Note the similitude of this expression with the first one in Eq. (8). However, in contrast to the Kretschmann and Otto layouts, these plasmonic microcavities can exhibit high transmittance over a broad range of thicknesses. This is related to the amplitude transmission coefficients being greater than one, as can be seen in Fig. 4(b). While $d \leq max(D)$), the reflectance and transmittance of the whole system, remain smaller than one and are limited by absorption in the metallic mirror layers. From there, the transmission peak decreases continuously until the opaque limit is achieved.

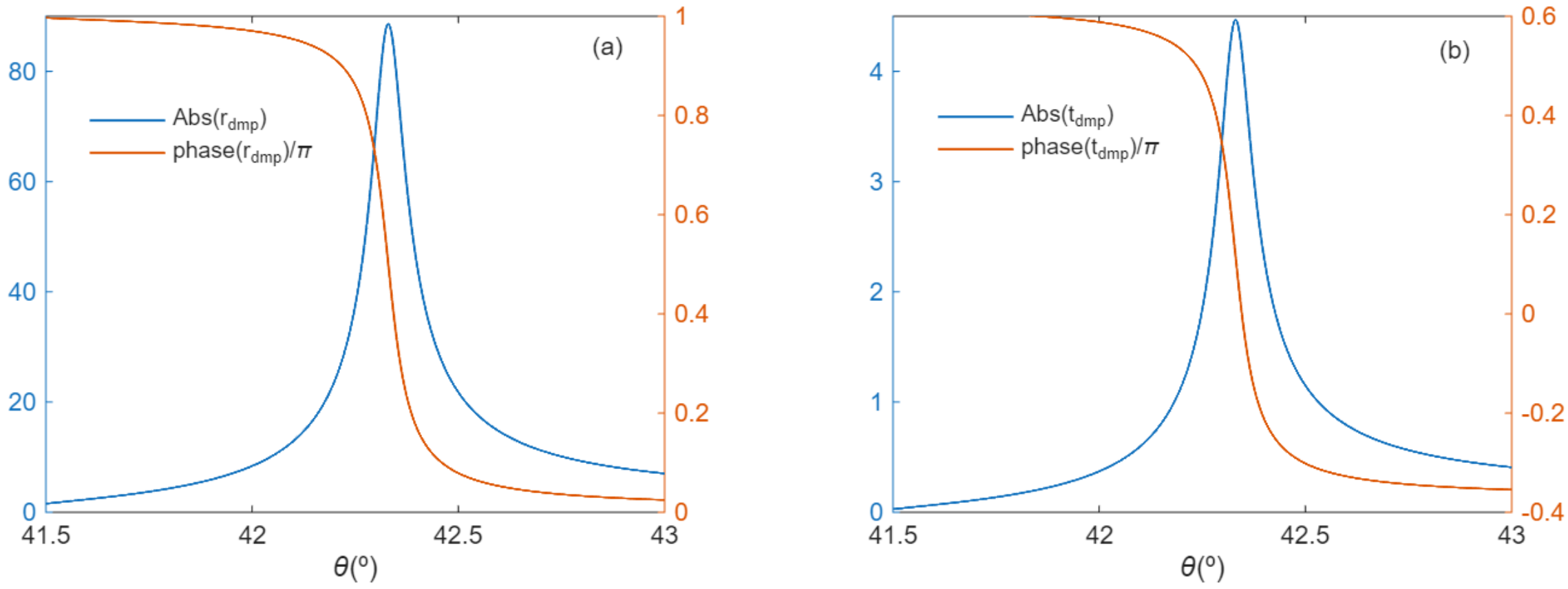

Fig. 4. (a) shape of $r_{dmp}$ as a function of the angle of incidence in the prism. The prism material is BK7, the mirror is a thin Ag layer with $w$ = 50 nm, and the intracavity medium is air. The wavelength is λ = 800 nm. (b) shape of the corresponding $t_{dmp}$ coefficient. It can be shown that $t_{pmd} = t_{dmp} \, k_{pn}/k_{dn}$.

The present analysis refers to angles of incidence greater than the critical angle $\theta_{cr}$, as previously defined, where the waves in the microcavity are evanescent. It connects the Fabry–Pérot regime, for which $\theta < \theta_{cr}$, and $||r_{dmp}| \leq 1$ and the waves are harmonic, with the waveguide regime, in which waves are guided in the thin metal film with evanescent tails in dielectric surroundings, at CSP resonances. These modes correspond to the poles of $r_{dmp}$.

Unusual reflection coefficients are not limited to plasmonic resonances, but can also appear in purely dielectric structures. In fact, whenever a reflection coefficient has poles, it may take arbitrarily large values in the vicinity of those poles. As a simple example, we consider a three-medium structure denoted by the indices 1, 2 and 3, where the central medium is a thin high-refractive-index layer with $n_2 > n_1, n_3$; see Fig. 1(d). Both TE and TM waves can be considered. This structure, consisting of a slab layer with a refractive index higher than that of the surrounding media, corresponds to a step-index waveguide when the propagation constant lies in the interval $k_0 \max(n_1, n_3) \leq \beta \leq k_0 n_2$. In this interval, the wave is propagating in the core layer and evanescent in the outer media. Therefore, the Fresnel reflection coefficients at the core boundaries have unit modulus and can be written as: $r_{2j} = e^{i\delta_{2j}},\ j = 1{,}3$. That means than for a core layer of thickness *d* the 1 to 3 reflection coefficient is:

$$r_{123} = \frac{-e^{i\delta_{21}} + e^{i\delta_{23}} e^{i2k_{2n}d}}{1 - e^{i\delta_{21}} e^{i\delta_{23}} e^{i2k_{2n}d}} = -\frac{\sin(k_{2n}d + \delta_{23}/2 - \delta_{21}/2)}{\sin(k_{2n}d + \delta_{23}/2 + \delta_{21}/2)}$$
$$= \sin\delta_{21} \cot(k_{2n}d + \delta_{23}/2 + \delta_{21}/2) - \cos\delta_{21}$$

(11)

A similar expression is obtained for the $r_{321}$ coefficient by interchanging $\delta_{21}$ and $\delta_{23}$. In this equation, $k_{2n} = \sqrt{\varepsilon_2 k_0^2 - \beta^2}$ is the component of the wavevector in the core layer normal to the interfaces. Since both coefficients are quotients of sine function, their values are not bounded. For example, Fig. 5(a) shows $r_{123}$ as a function of $N = \beta/k_0$ for an air - BK7 glass–$SiO_2$ structure d = 5 μm. The discontinuities correspond to the poles of $r_{123}$. They fulfill:

$$2k_{2n}d + \delta_{21} + \delta_{23} = 2m\pi \quad m \in \mathbb{Z}^+$$

(12)

This condition can also be written in complex notation as:

$$r_{21} r_{23} e^{i2k_{2n}d} = 1$$

(13)

Relations (12) and (13) are the resonance conditions for the guided modes of the step-index dielectric waveguide with constant of propagation $\beta$. Therefore, we again find an equivalence between modes and poles of the corresponding amplitude reflection coefficients. Naturally, the corresponding amplitude transmission coefficients have the same poles.

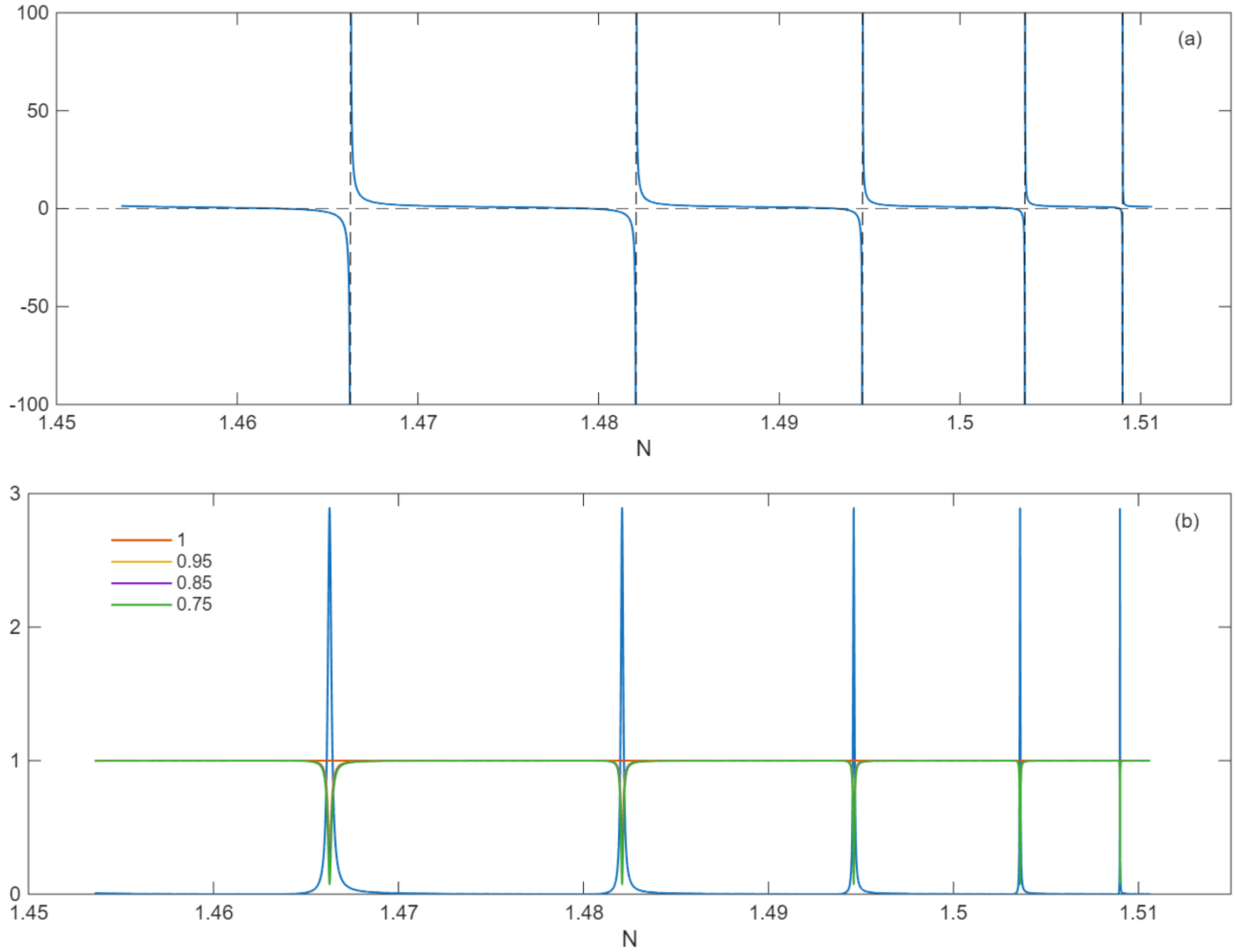

Fig. 5. (a) $r_{123}$ versus $N = \beta/k_0$, where $\beta$ is the propagation constant. We observe discontinuities at the poles of $r_{123}$. (b) $|r_{0/3}|$ (blue line) and $|r_{0123}|$ for $\alpha = 1, 0.95, 0.85, 0.75$. The configuration is as follows: $SF_{11}$ (prism), air w = 200 nm (cladding), $BK_7$ d = 5 μm (core), $SiO_2$ substate. The wavelength Is λ = 800 nm.

The reflection coefficient $r_{123}$ appears in the modelling of step index waveguides prism coupling device and in the m-line technique for characterizing thin films [15], see Fig. 1(e). Denoting the high-refractive-index prism by the index 0, the reflection coefficient of the whole system is

$$r_{0123} = r_{01} + r_{0/3}; \quad r_{0/3} = \frac{t_{01}t_{10}r_{123}e^{i2k_{1n}w}}{1 - r_{10}r_{123}e^{i2k_{1n}w}} \tag{15}$$

where $w$ is the thickness of layer 1, which is typically an air gap. In this system, typically, $n_0 > n_2 > n_3 > n_1$. The coefficient $r_{123}$ is as in eq. (11) while $r_{01}$ is a complex coefficient with unit modulus in the region of interest. Furthermore, in this region $k_{1n} = \sqrt{\varepsilon_2 k_0^2 - \beta^2} = i\sqrt{\beta^2 - \varepsilon_2 k_0^2}$. As in the SPR prism-coupling configurations discussed above, the reflected wave results from the superposition of two contributions with a similar physical interpretation: a direct reflection at the prism–gap interface and a contribution associated with light that penetrates into the multilayer structure and is reflected back. In Fig. 5(b) we plot $|r_{0/3}|$, which shows pronounced peaks close to the poles of $r_{123}$. Despite these peaks, the reflectance, $|r_{0123}|^2$ is equal to one for any incidence condition above the critical angle, owing to total internal reflection. In the present case, we refer to this process as extended total reflection, since total reflection does not occur only at the first interface. Instead, the light penetrates into the system, extends through media 1 and 2, reaches medium 3, and is finally reflected back to the prism. This ideal behaviour with a reflectance of unity relies on perfect translational symmetry. In practice, however, the system may present inhomogeneities associated with light scattering, surface roughness, or interface distortions, among other effects, which break this symmetry. Although these perturbations may modify the term $r_{0/3}$ in several ways, we can model their effect on the reflection in a simple phenomenological way by introducing a real coefficient $\alpha < 1$, so that

$$r_{0123} = r_{01} + \alpha r_{0/3} \tag{16}$$

This modification reduces the contribution associated with the resonant term $r_{0/3}$. As a result, reflectance dips appear close to the peaks of $|r_{0/3}|$, provided that the thickness w of the first layer remains small. In Fig. 5(b), we show the reflectance for several values of α. In general:

$$R = |r_{0123}|^2 = 1 - \alpha(1-\alpha)|r_{0/3}|^2 \tag{17}$$

As α decreases, the reflectance dips become deeper, reaching their minimum depth around α = 0.5. In practice, this coefficient is expected to depend on both the angle of incidence and the wavelength.

In summary, we analysed several optical configurations in which resonant excitation occurs through evanescent or damped waves. Although these systems are physically different, they share a common feature: some field-amplitude reflection or transmission coefficients can exceed unity by a large amount, or even become unbounded. These large coefficients are associated with the proximity of resonances, and in the limiting case with the poles corresponding to guided modes. However, this behaviour does not imply a violation of energy conservation, since the unusually large coefficients cannot be directly measured as reflectance or transmittance. Rather, they arise from the formal extension of conventional reflection and transmission coefficients to situations involving evanescent or damped waves. In this sense, they should be understood as relations between different components of the electromagnetic field rather than as direct ratios of the optical power carried by independent propagating waves. The examples discussed here demonstrate that such coefficients naturally arise from the application of boundary conditions in resonant multilayer systems. This interpretation provides a unified framework for understanding large amplitude coefficients in plasmonic interfaces, prism-coupling configurations, microcavities and dielectric waveguides. The same reasoning can be extended to other optical structures, including more complex plasmonic systems, dielectric waveguides, grating and waveguide couplers, metamaterials and other resonant photonic devices.

Funding

This research was funded through projects by the Spanish Ministry of Science, Innovation and Universities (PID2024-156552OA-I00), Universidade de Santiago de Compostela (USC 2024-PU031) and Xunta de Galicia (GRC ED431C 2024/06).

Data Availability Statement

The data that support the findings of this study are available from the corresponding author upon reasonable request.

Conflicts of interest

The authors declare that they have no known competing financial interests or personal relationships that could have appeared to influence the work reported in this paper.